\documentclass[showpacs,aps,twocolumn]{revtex4}
\usepackage{graphicx}

\begin{document}

\title{Current and vortex statistics in microwave billiards}
\author{Michael Barth}
  \email{michael.barth@physik.uni-marburg.de}
\author{Hans-J\"urgen St\"ockmann}
  \email{stoeckmann@physik.uni-marburg.de}
\affiliation{Fachbereich Physik, Philipps-Universit\"at Marburg, Renthof 5, D-35032 Marburg, Germany}
\date{\today}

\begin{abstract}
Using the one-to-one correspondence between the Poynting vector in a microwave billiard
and the probability current density in the corresponding quantum system
probability densities and currents were studied in a microwave billiard with a ferrite
insert as well as in an open billiard. Distribution functions were obtained for probability densities, currents and vorticities. In addition the vortex pair correlation function could be extracted. 
For all studied quantities a complete agreement with the predictions from the random-superposition-of-plane-waves approach was obtained.
\end{abstract}


\maketitle

\section{Introduction}

A particularly successful approach to describe the statistical properties of the
eigenfunctions of chaotic billiards, dating back to Berry \cite{Ber77a}, assumes that at
any point not too close to the boundary the wave function can be described by a random
superposition of plane waves,
\begin{equation}\label{eq:superposition}
\psi(r) = \sum_n a_n e^{\imath k_n r}\,,
\end{equation}
where the modulus $k=|k_n|$ of the incoming waves is fixed, but directions $k_n/k$ and
amplitudes $a_n$ are considered as random. In billiards with time-reversal symmetry there is in addition the restriction that the wave function has to be real. This ansatz cannot
be strictly true. It completely ignores the boundary conditions at the billiard walls,
but this is of no importance as long as the wave length is small compared to the billiard
size.

As a consequence one expects a Gaussian distribution for the wave function amplitudes
$\psi$, or, equivalently, a Porter-Thomas distribution for their squares $|\psi|^2$. For
the spatial auto-correlation function 
$c(r)=\left\langle\psi^*(\bar{r}+r)\psi(\bar{r})\right\rangle_{\bar{r}}$ a Bessel function is found. 
These predictions were first verified by McDonald and Kaufman in their disseminating 
papers on stadium wave functions \cite{McD79,McD88}. It is impossible to mention all 
works which have been published hitherto on this subject. As another consequence of 
ansatz (\ref{eq:superposition}) the gradient of the wave function is Gaussian distributed, too, and
uncorrelated to the wave function. This has been used to calculate the distribution of eigenvalue
velocities and the velocity auto-correlation function for the case of a local parameter
variation \cite{Bar99c}.

The approach is not restricted to quantum-mechanical systems, or systems where there is a
one-to-one correspondence to quantum mechanics such as quasi-two-dimensional microwave
billiards. Porter-Thomas distributions are found as well in the squared amplitudes of
vibrating plates \cite{Sch97}. In an experiment on a three-dimensional Sinai microwave
billiard, having no quantum mechanical equivalent, the electromagnetic field
distributions as well as their spatial correlations could be explained assuming a random
superposition of plane electromagnetic fields \cite{Doer98b,Eck99}. In a very recent
experiment on light propagation through distorted wave guides, finally, the observed
transversal field patterns could be described again by ansatz (\ref{eq:superposition}) \cite{Doy02}. All
these examples show that the random-superposition-of-plane-waves approach is not of a
quantum-mechanical origin, but holds for all types of waves.

If the billiard is opened, or if time-reversal symmetry is broken, the wave function
acquires an imaginary part, 
\[
\psi = \psi_R + \imath\,\psi_I,
\]
 with the consequence that in dependence
of the relative fractions of real and imaginary parts the distribution of $|\psi|^2$ changes from
Porter-Thomas to single exponential behavior. An explicit formula describing the
distribution in the transition regime has been given by different authors
\cite{Zyc91,Kan96,Pni96,Sai02}. The same function has been derived by \v{S}eba {\it et al.} for the distribution of scattering matrix elements in a partially opened
microwave billiard \cite{Seb97}. Wu {\it et al.} studied amplitude distributions and spatial
auto-correlation functions in a microwave billiard with one ferrite-coated wall to break
time-reversal symmetry, and found quantitative agreement with the results expected from
the random-superposition-of-plane-waves approach \cite{Wu98}. In a recent paper by Ishio {\it et al.} also deviations of this formula due to scars and in regular systems are studied \cite{Ish01}.

More recently the interest focussed on the current statistics in open systems. Saichev {\it et
al.} calculated the distribution of currents \cite{Sai02}. The properties of current
vortices have been studied by Berry and Dennis, who gave analytic expressions for
different types of vortex spatial auto-correlation functions \cite{Ber00b}. Independently, such auto-correlation functions as well as the distribution of nearest distances between nodal points have been studied in Ref.~\cite{Sai01}.

\begin{figure*}
\includegraphics[width=5.9cm]{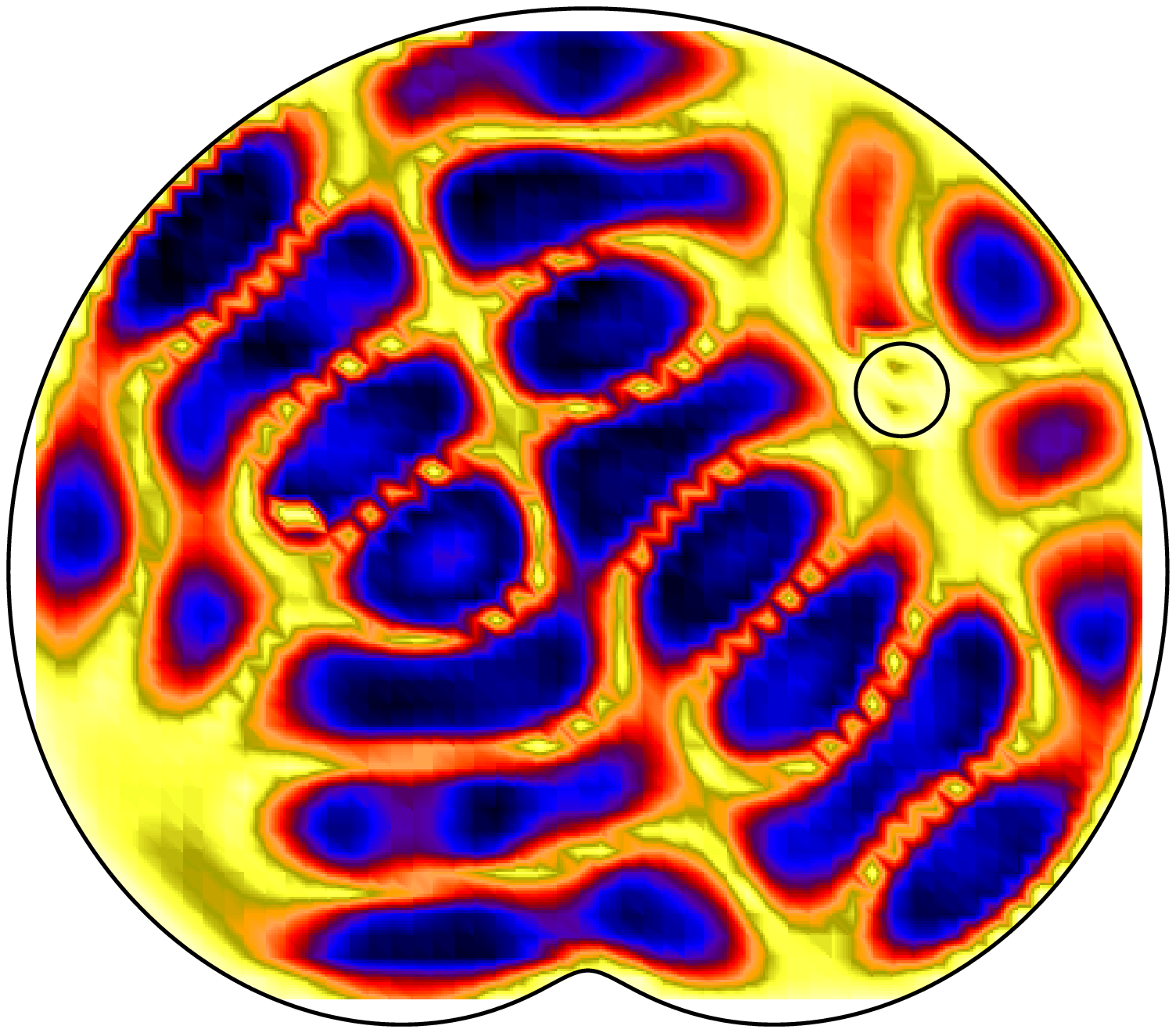}
\includegraphics[width=5.9cm]{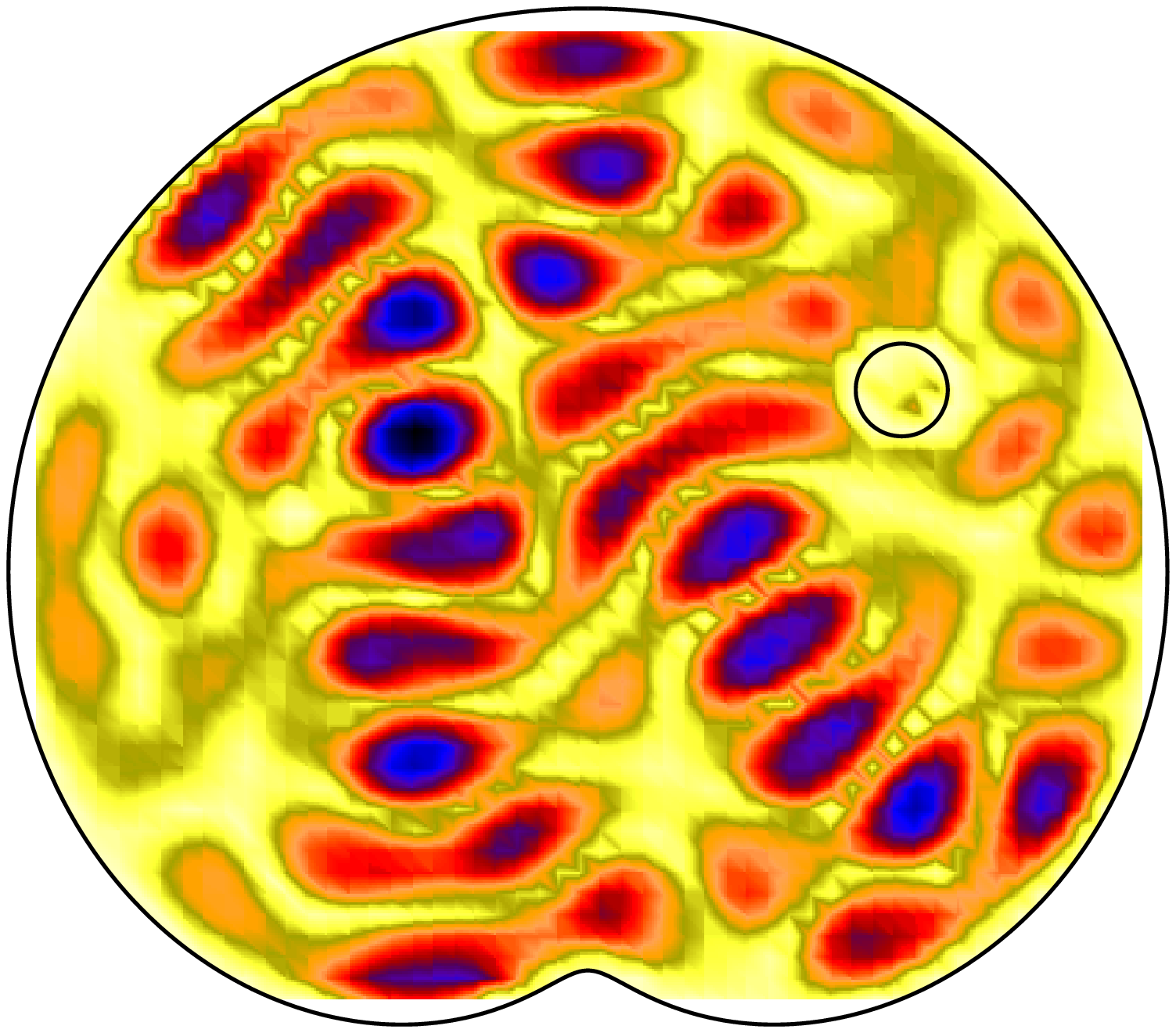}
\includegraphics*[width=5.9cm]{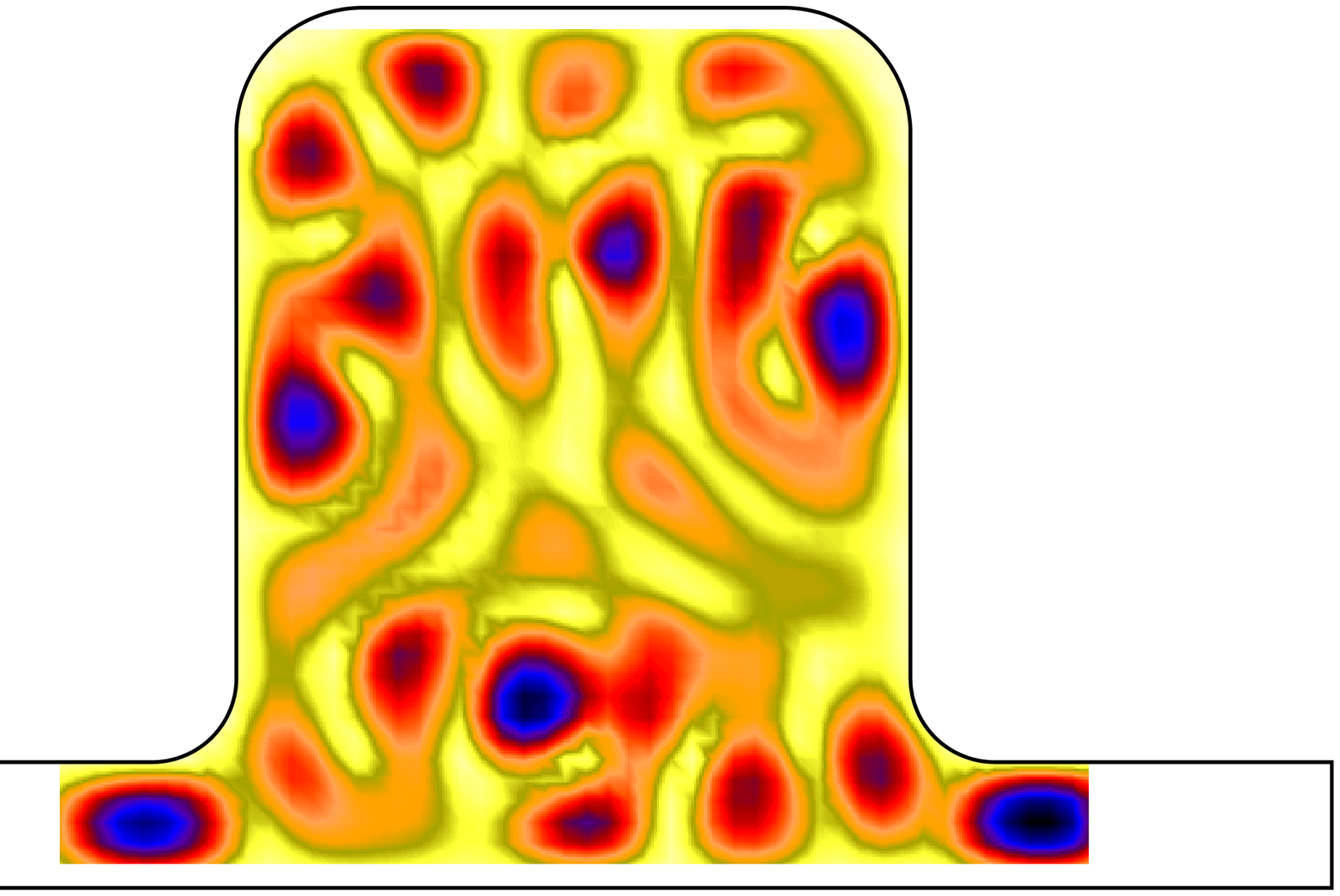} \\
\includegraphics[width=5.9cm]{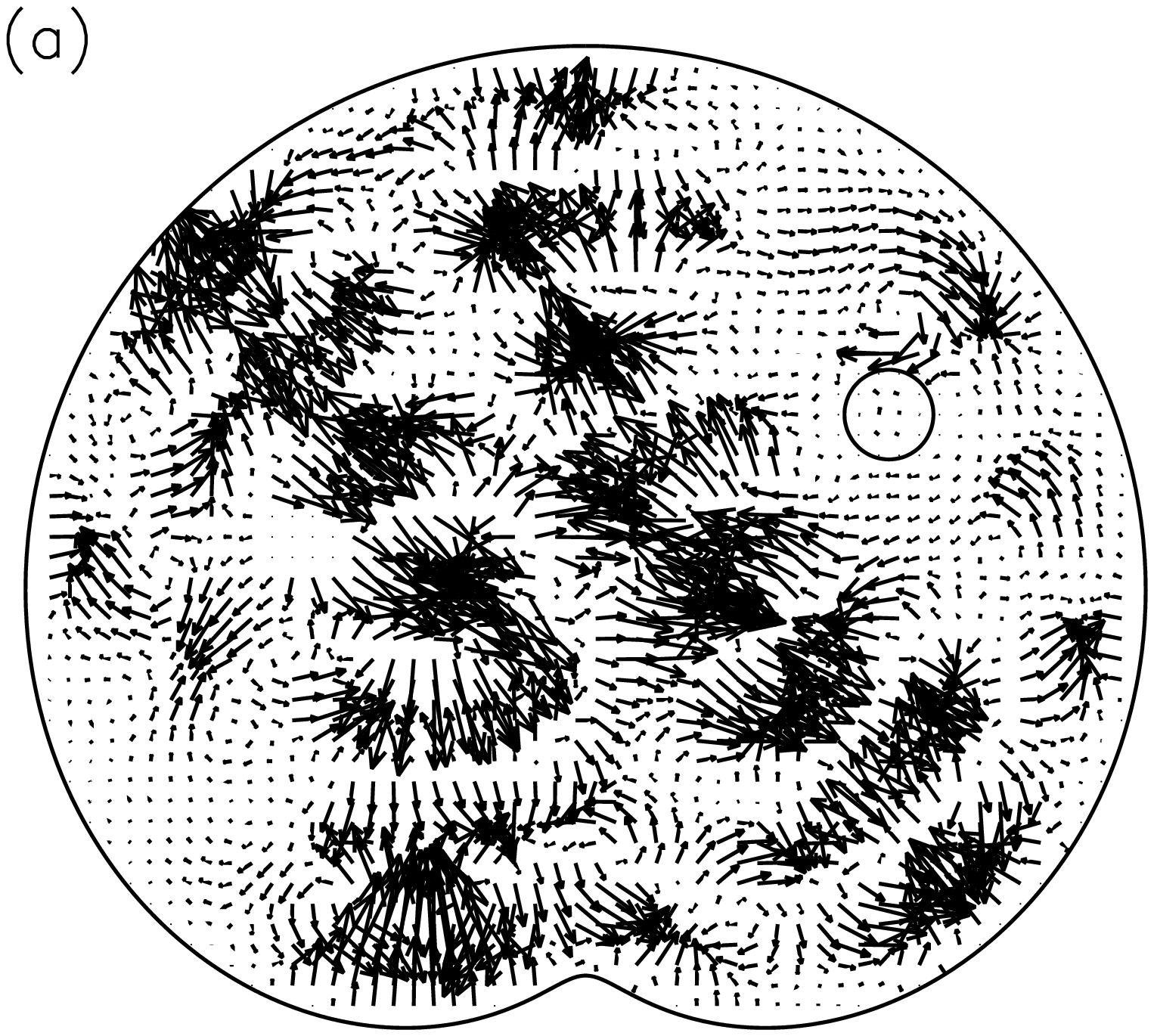} 
\includegraphics[width=5.9cm]{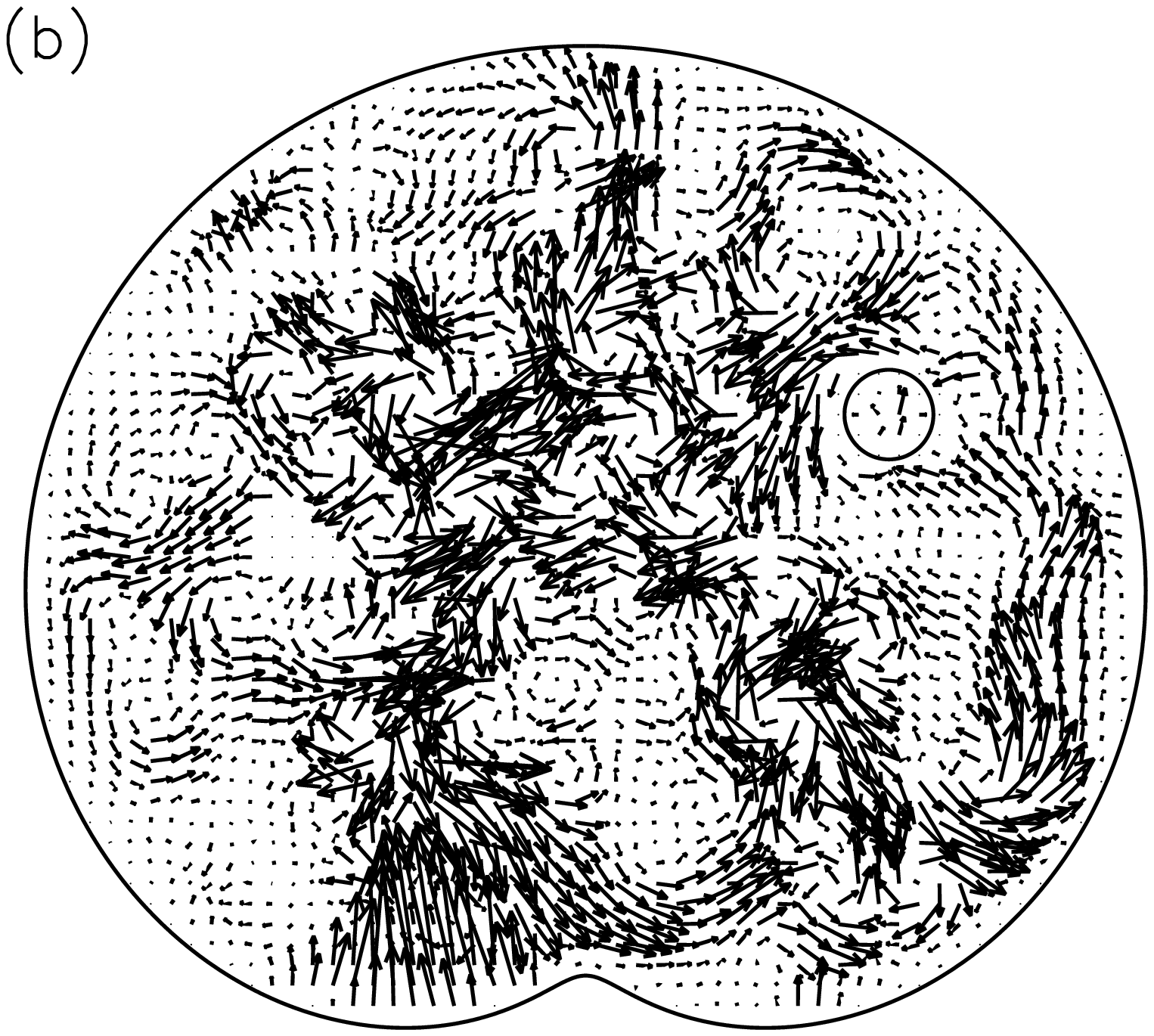} 
\includegraphics*[width=5.9cm]{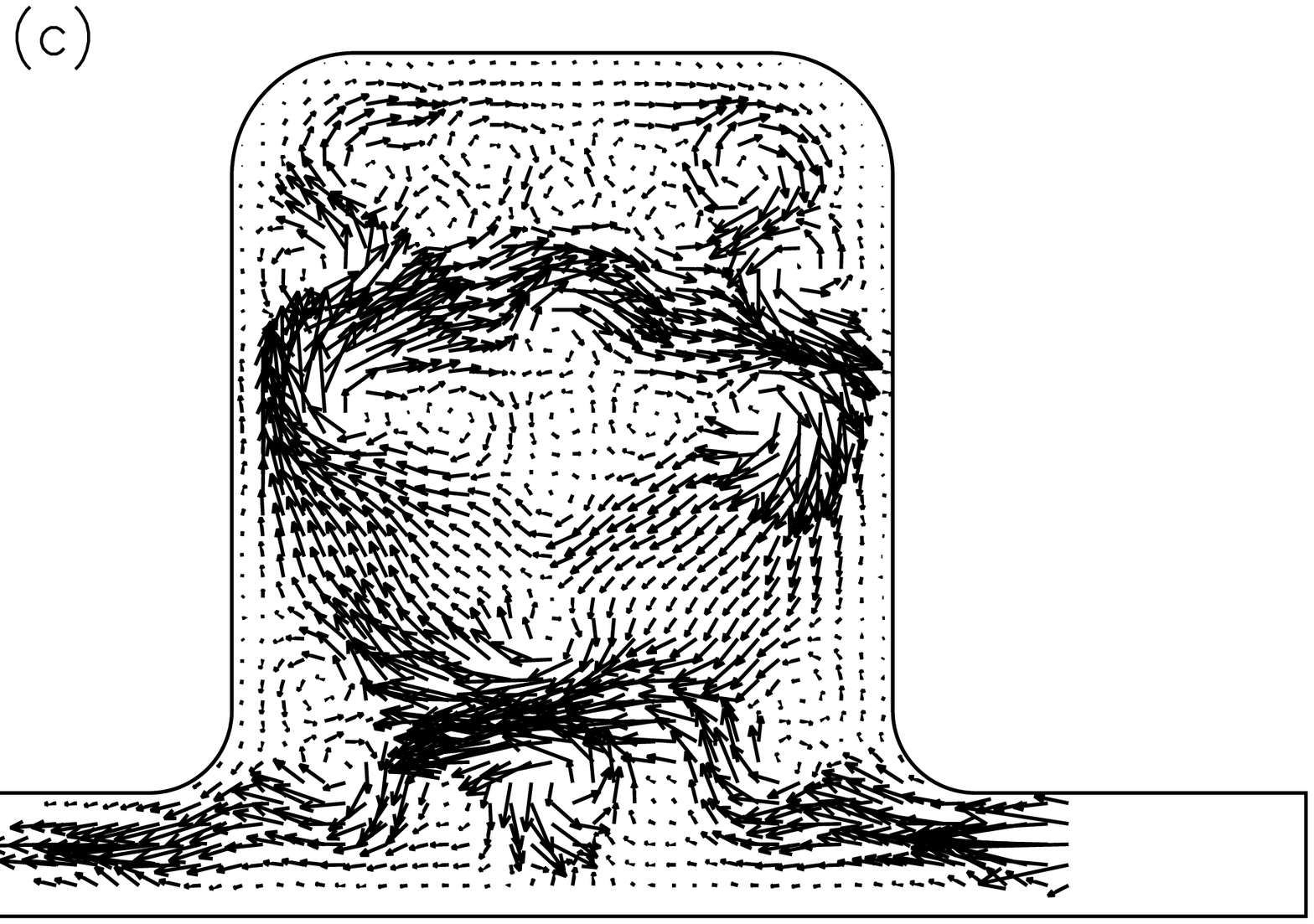}
\caption{Map of $|\psi|^2$ (top) and of the current (bottom) in the ferrite billiard at 5.19~GHz (a) and 6.41~GHz (b), and for the open billiard at 5.77~GHz (c). 
The intensity has been converted into a gray scale with the intensity
increasing from white to black. The lengths of the arrows in the current map
correspond to the magnitude of the Poynting vector.
For the field distribution in panel (a) there is a strong flow into the probing antenna at points of high intensity. In the two other examples the flow into the probing antenna is negligible.}
\label{fig:billiards}
\end{figure*}

Experimental results on current statistics and the distribution of nodal points have not
been available as yet. The method of choice to study such questions are microwave
techniques. In a previous work of our group we could demonstrate that currents can be
easily obtained in an open microwave billiard \cite{Seb99}. In the present paper we
show results on current distributions in two types of billiards,  
verifying a number of predictions given in the above mentioned
papers for the first time also in experiment.

\section{Experiment}

The basic principles of the experiment are described in detail in Ref.~\cite{Kuh00b}.
Therefore we concentrate on the aspects being of relevance to the present study. Two
different billiards have been used, both resonators have a size of about 25~cm and a height of 8~mm. 
One of them is an open system of a rectangular shape with rounded corners and two openings on opposite sides containing entrance and exit antenna (see Fig.~\ref{fig:billiards}c).
A third movable antenna was used to map the field distribution in the resonator on a square grid
of period 5~mm. The same system was used for a quantum-dot analog study previously \cite{Kim02}.
The two fixed antennas have a metallic core of diameter 1~mm, and a teflon coating for stronger coupling; the movable antenna was a thin wire of diameter 0.2~mm. The lenghts of all antennas were about 6~mm. 
The quantity directly accessible in the experiment is the scattering matrix \cite{Ste95}. From a reflection measurement as a function of the antenna position a mapping of the modulus of the wave function can be obtained; to get the sign as well, the transmission between two antennas is needed. By the presence of the antennas the resonances are somewhat broadened and shifted, but in chaotic systems this does not change the universal distributions. (For details see Ref.~\cite{Kuh00b}.)

The second system studied is a Robnik, or lima\c{c}on billiard \cite{Rob83,Rob84b}. It  can be obtained by a 
complex mapping of the unit circle in the complex plain by means of the function 
$w=z+\lambda z^2$. The deformation parameter used in the experiment was $\lambda=0.4$. 
For this parameter the billiard is believed to be completely chaotic. (There are recent 
theoretical results indicating that this cannot be taken for sure \cite{Dul01}. But if there are still stable islands in the phase space, their corresponding volume will be very small.) A ferrite ring is introduced to 
break time-reversal symmetry (shape and position are included in Figs. \ref{fig:billiards}a and b). Ferrites have been used already repeatedly for this purpose \cite{So95,Sto95b,Wu98,Sch01d}. A detailed description of the function principle of the ferrites will be given in a forthcoming publication \cite{SchC}. During 
reflection at the ferrite the microwaves experience a phase shift with the
consequence that there will be currents through the billiard though it is completely
closed. These currents are a complete analogue to the persistent currents observed in mesoscopic structures \cite{Lev90}, for a recent review see Ref.~\cite{Moh99}. 
They will be the subject of a separate publication \cite{Vra}. In addition there will be currents due to the fact that the ferrite introduces considerable absorption into the system. In the present context it is not of relevance whether the currents are due to a break of time reversal symmetry or due to absorption.

It was demonstrated in Ref.~\cite{Seb99} that in quasi-two-dimensional billiards the
Poynting vector $\vec{S}=\frac{c}{4\pi} \vec{E}\times\vec{H}$ can be written as
\[
\vec{S} = \frac{c}{8\pi k} \mathrm{\, Im}\left[ E^*(r) \nabla E(r) \right],
\]
showing that there is a one-to-one correspondence with the current density
\[
\vec{j} = \frac{\hbar}{m} \mathrm{\, Im}\left[ \psi^*(r) \nabla\psi (r) \right]
\]
in the corresponding quantum-mechanical system. A measurement of the electric field in
the resonator including the phase thus immediately yields the Poynting vector and by
means of the mentioned analogy the current density.

There is one problem with the experimental determination of field and current
distributions. The probe antenna moving through the billiard unavoidably gives rise to a
leakage current spoiling the statistical properties of the current distribution
\cite{Seb99}. 
The influence of the probe antenna is small as long as there is a strong flow
through the system. In this case the unavoidable leakage current into the probe
antenna is negligible. There are situations, however, where there is no or only
little flow, e.g. for the open dot, if the total transmission is close to zero,
or for the ferrite billiard, if there are strong standing waves present, as in
Fig.~\ref{fig:billiards}a. In such cases the leakage current is no longer negligible, and may
even become dominating.
There is only one way to avoid this problem: the
frequencies have to be chosen such that the overall amplitudes are moderate. As an
example Fig.~\ref{fig:billiards}a shows a mapping of $|\psi|^2$ and of the current distribution in the ferrite
billiard at a frequency where there is a strong flow into the measuring antenna at points
of high intensity. In Fig.~\ref{fig:billiards}b, on the other hand, the frequency has been chosen
such that there is no noticeable current into the probe antenna.

\section{Intensity distributions}

\begin{figure}
\includegraphics[width=8cm]{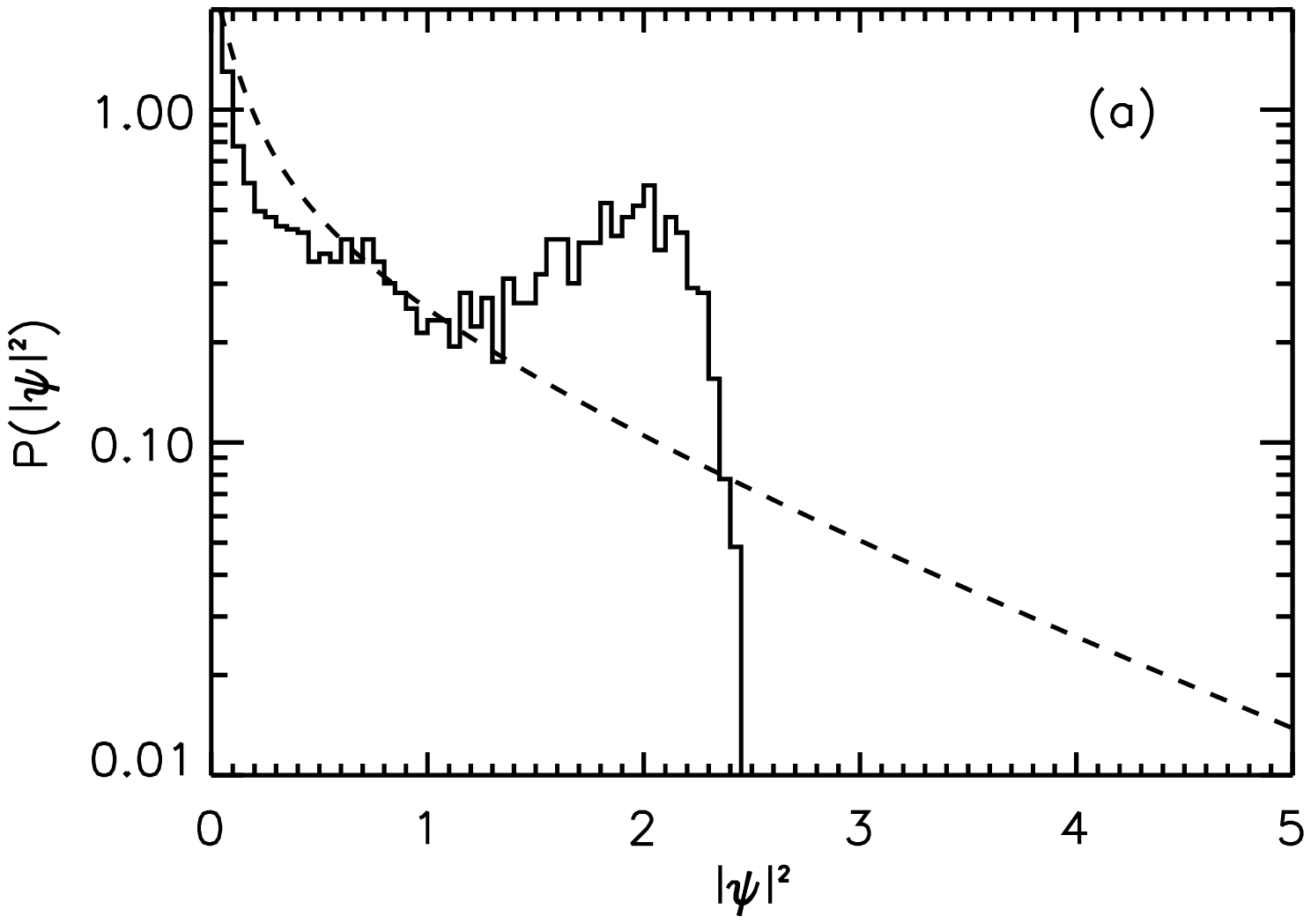}
\includegraphics[width=8cm]{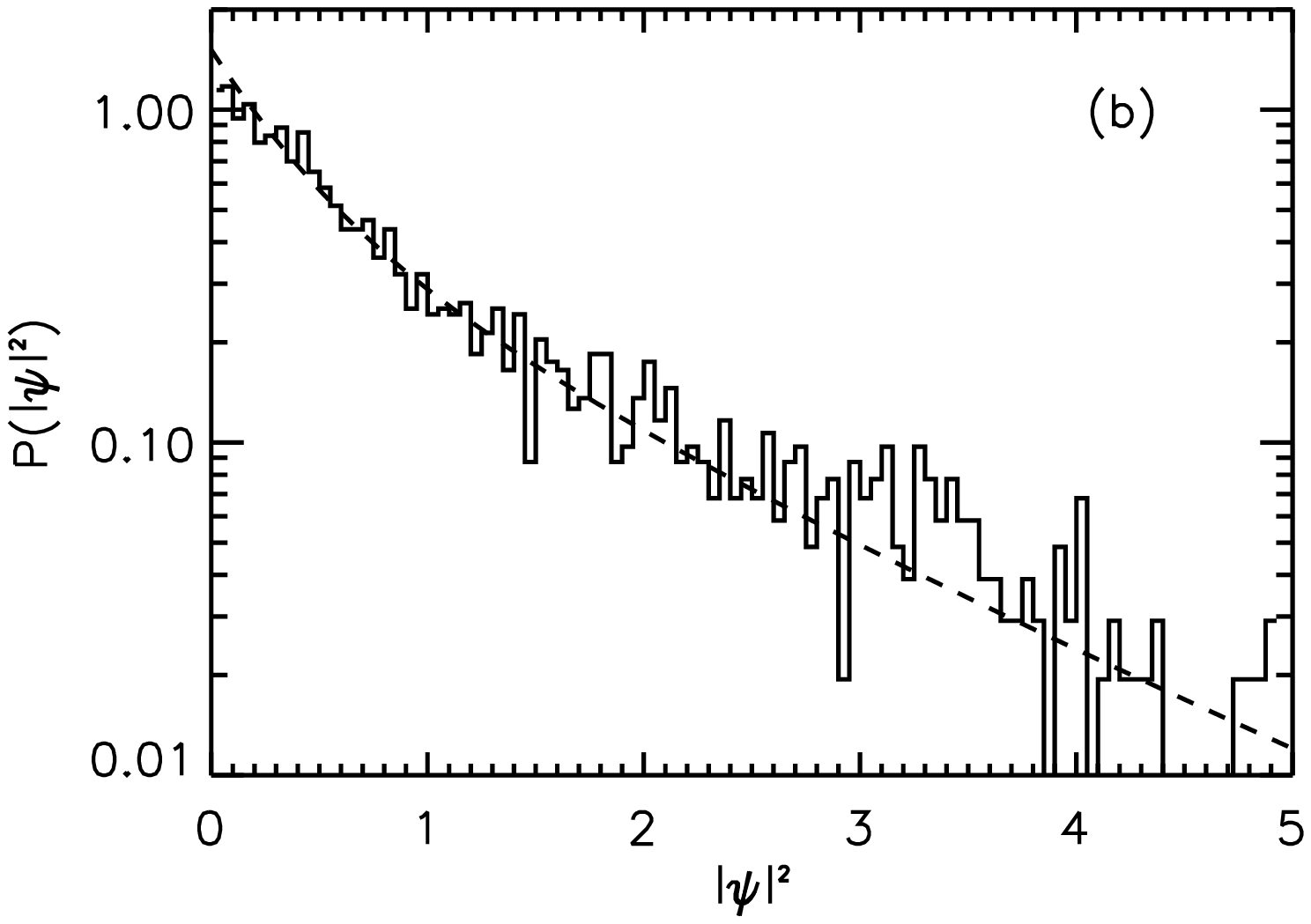}
\includegraphics[width=8cm]{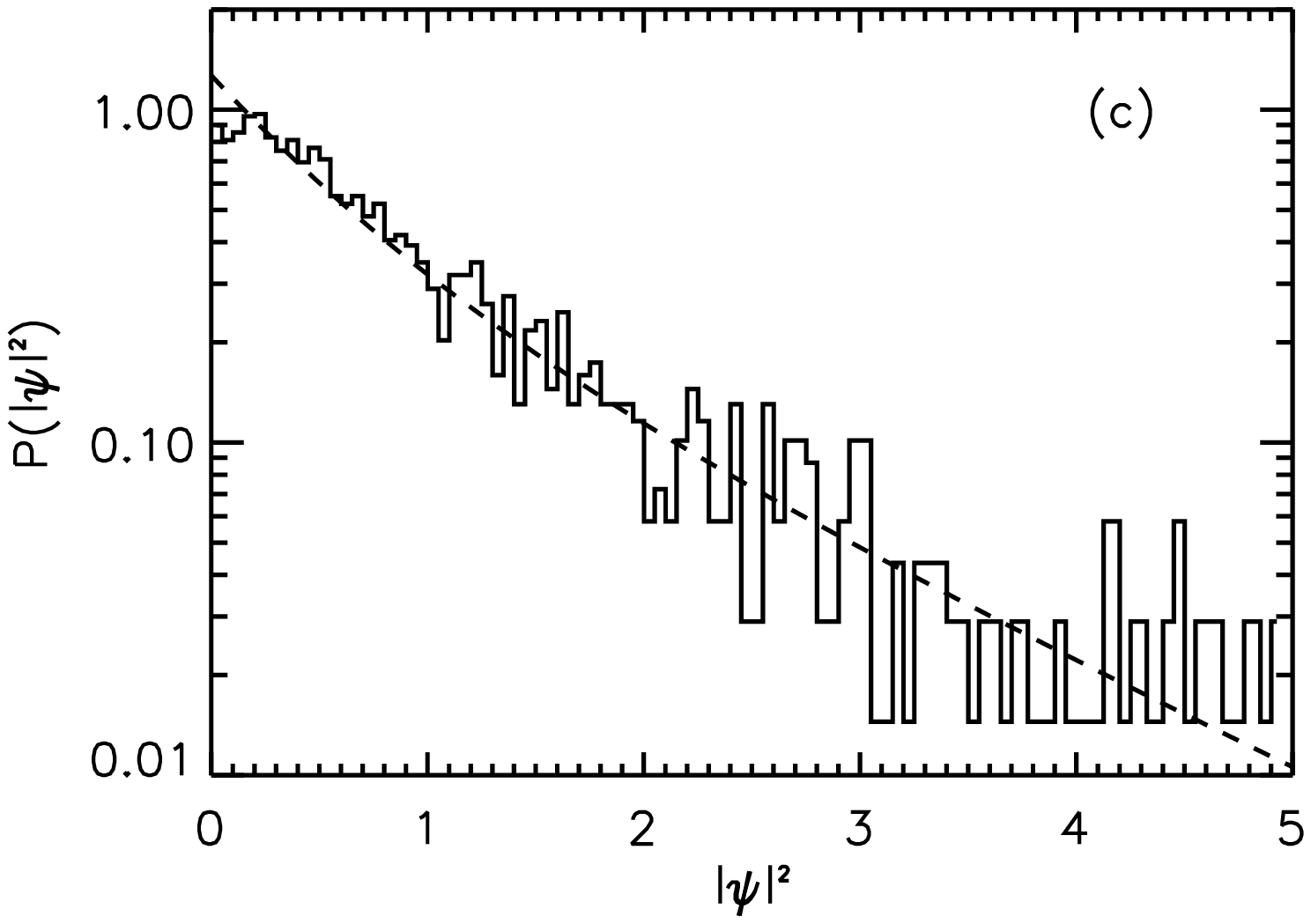}
\caption{Distributions of $\rho=|\psi|^2$ for  the ferrite billiard at 5.19~GHz (a), 6.41~GHz (b), and for the open billiard at 5.77~GHz (c), corresponding to Fig.~\ref{fig:billiards}.
The dashed lines have been calculated from Eq.~(\ref{eq:prho}).}
\label{fig:intensity}
\end{figure}

\begin{figure*}
\includegraphics[width=8cm]{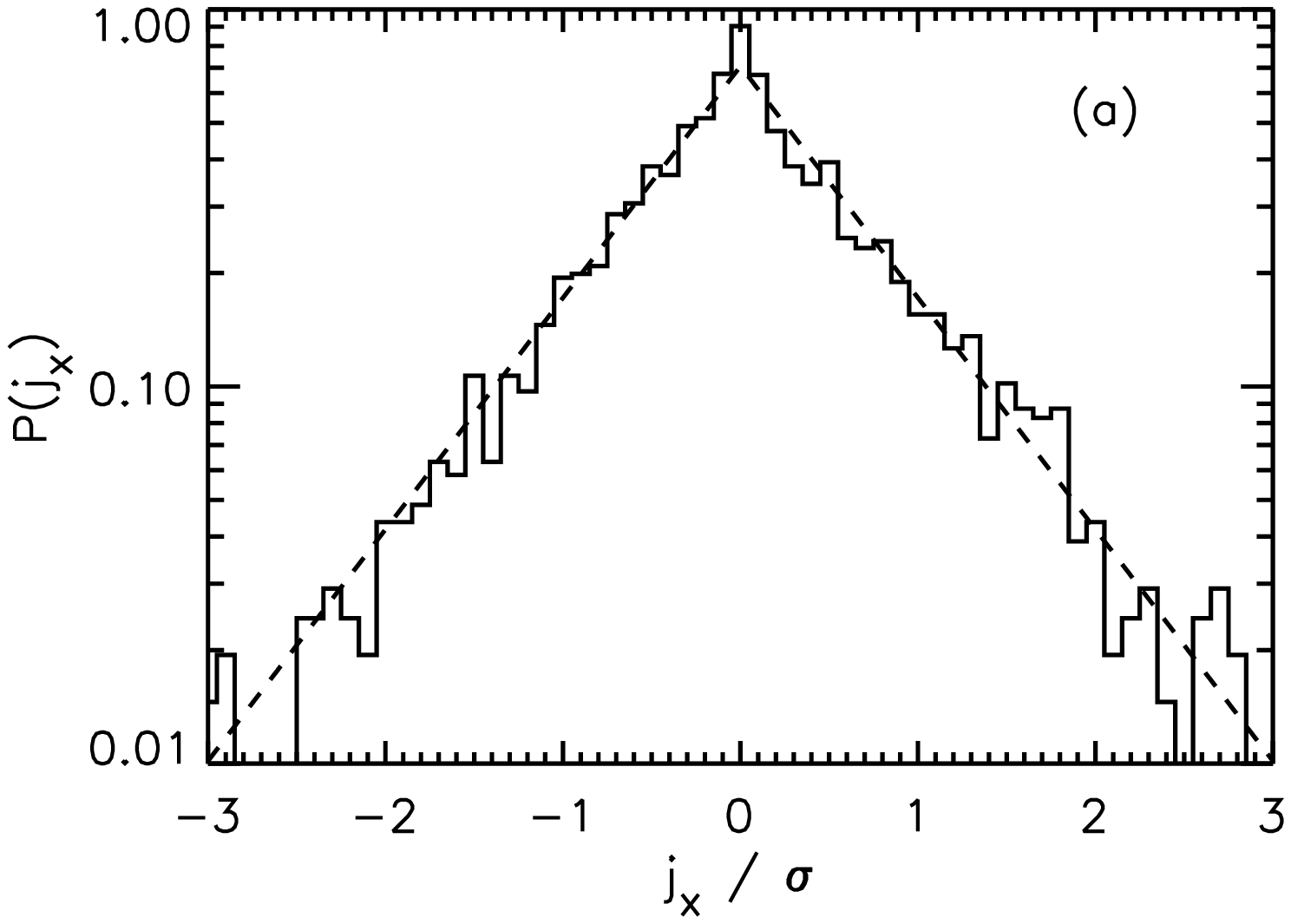}
\includegraphics[width=8cm]{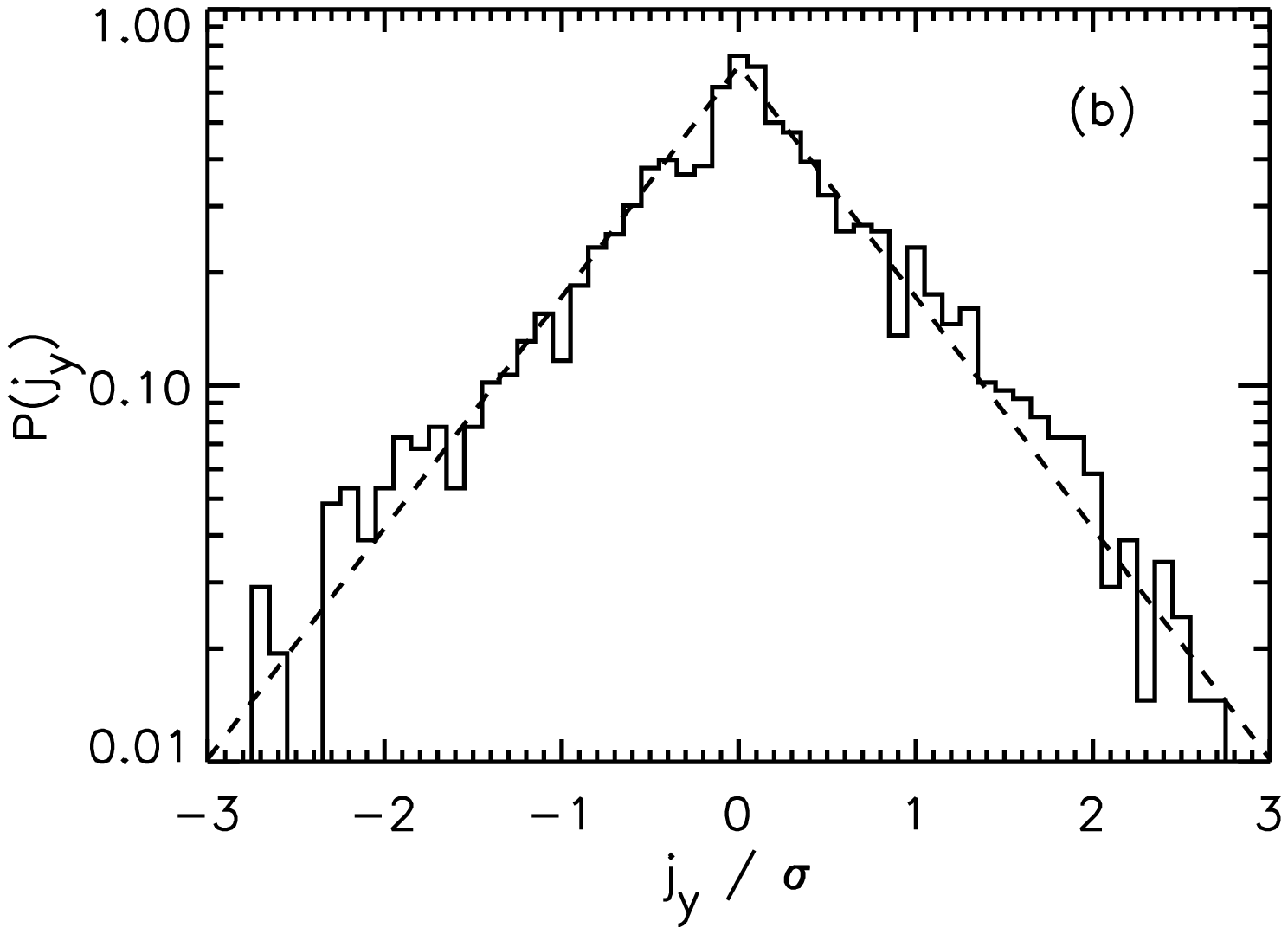}\\
\includegraphics[width=8cm]{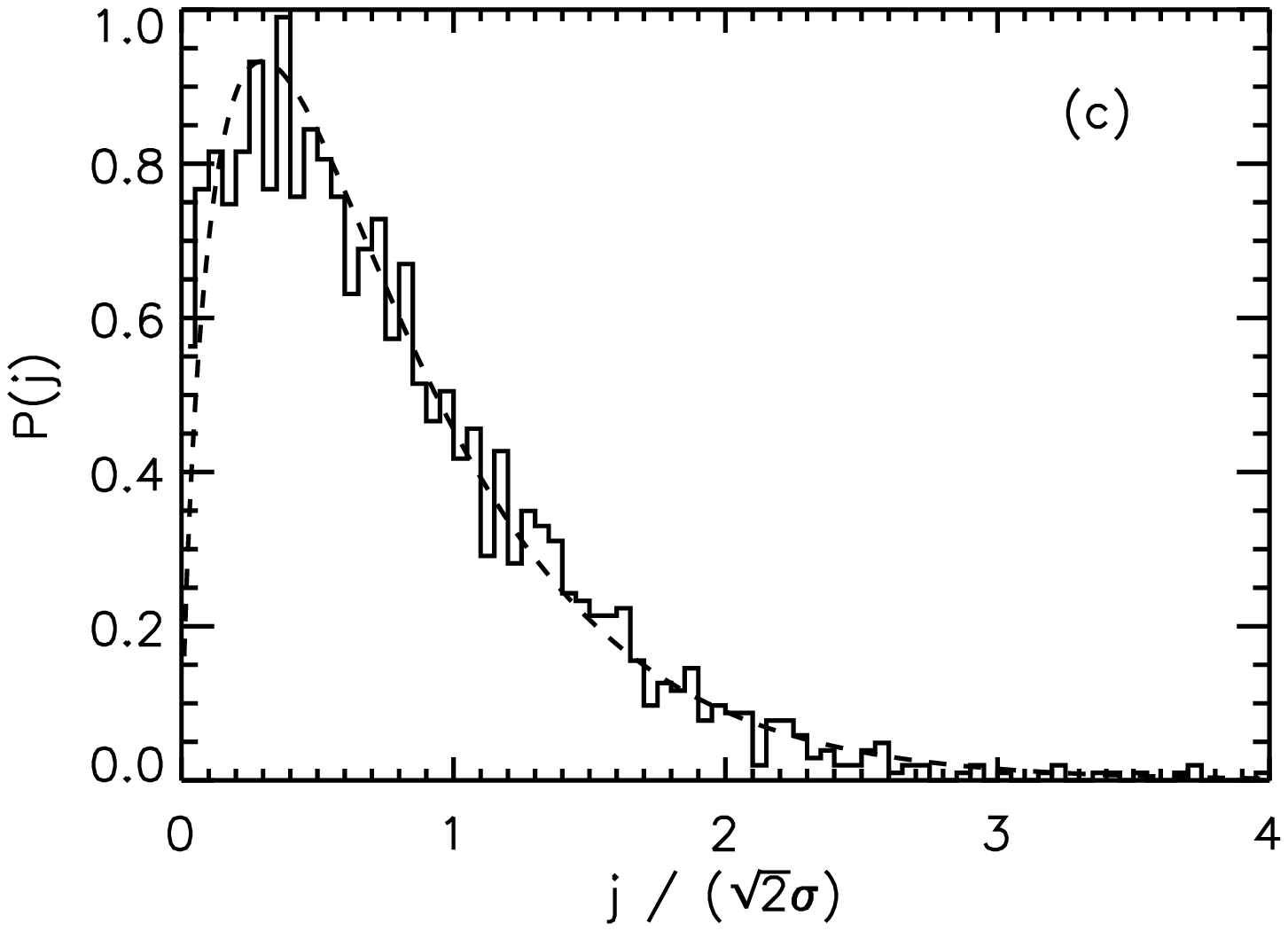}
\includegraphics[width=8cm]{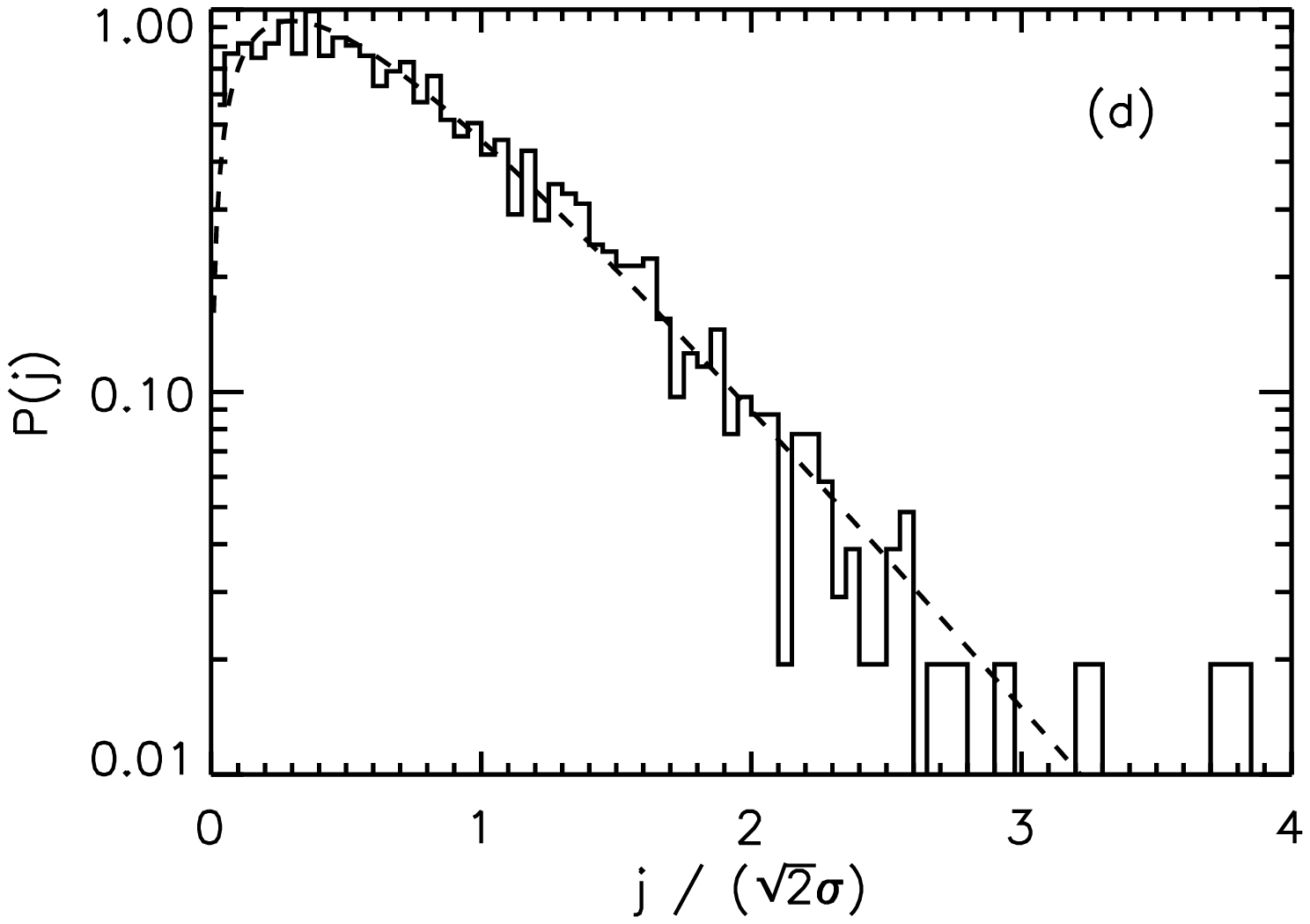}
\caption{Distributions of $j_x$ (a), $j_y$ (b), and $j$, the latter one both on a linear (c) and a logarithmic scale (d), for the ferrite billiard at 6.41~GHz, corresponding to Fig.~\ref{fig:billiards}b. 
The dashed lines are the theoretical expectations from Eqs. (\ref{eq:pjxy}) and (\ref{eq:pj}), respectively, where $\sigma=\sqrt{\langle j_{x,y}^2 \rangle}$, see Eq.~(\ref{eq:jparameter}).}
\label{fig:current_fb}
\end{figure*}

\begin{figure*}
\includegraphics[width=8cm]{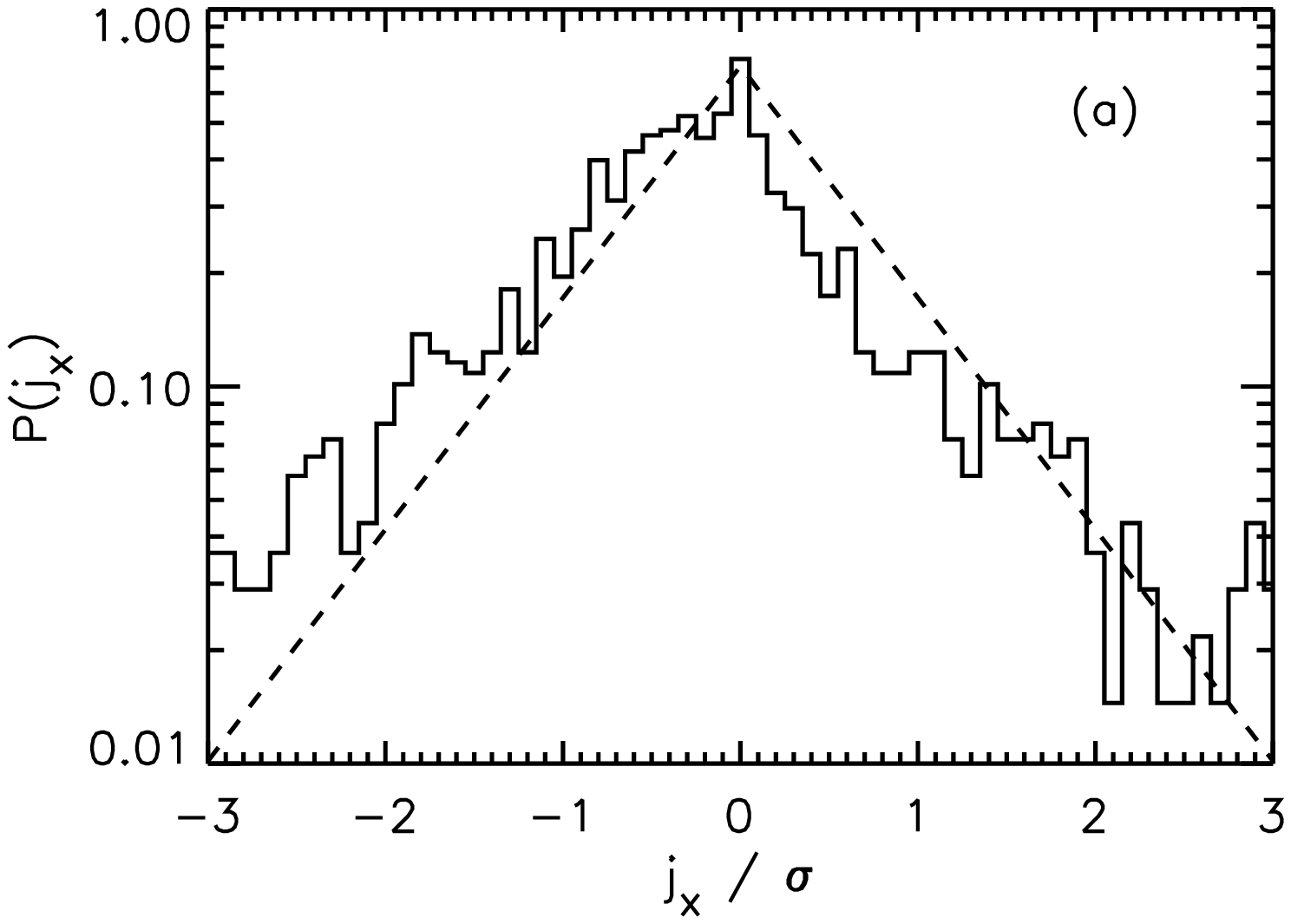}
\includegraphics[width=8cm]{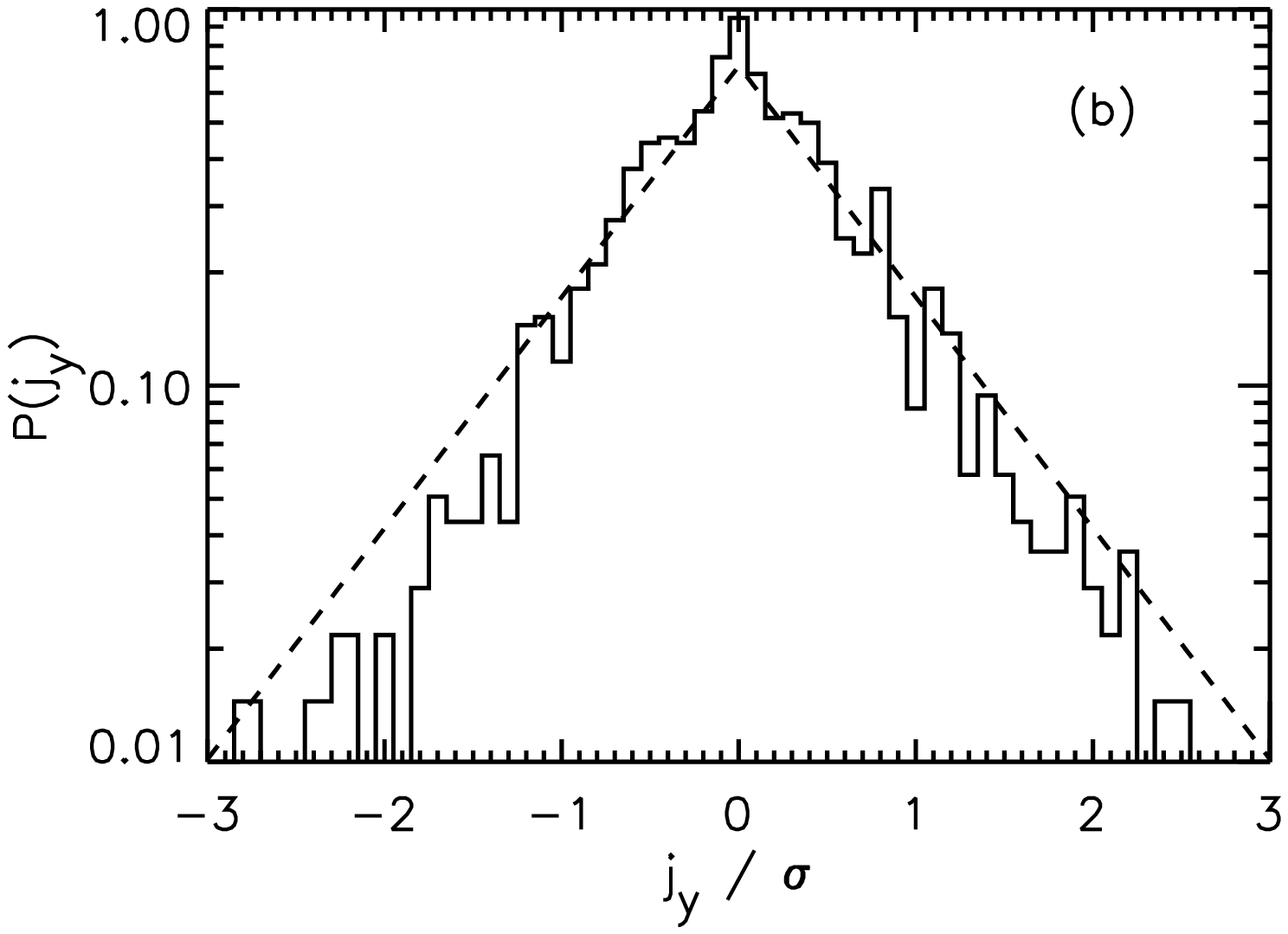}\\
\includegraphics[width=8cm]{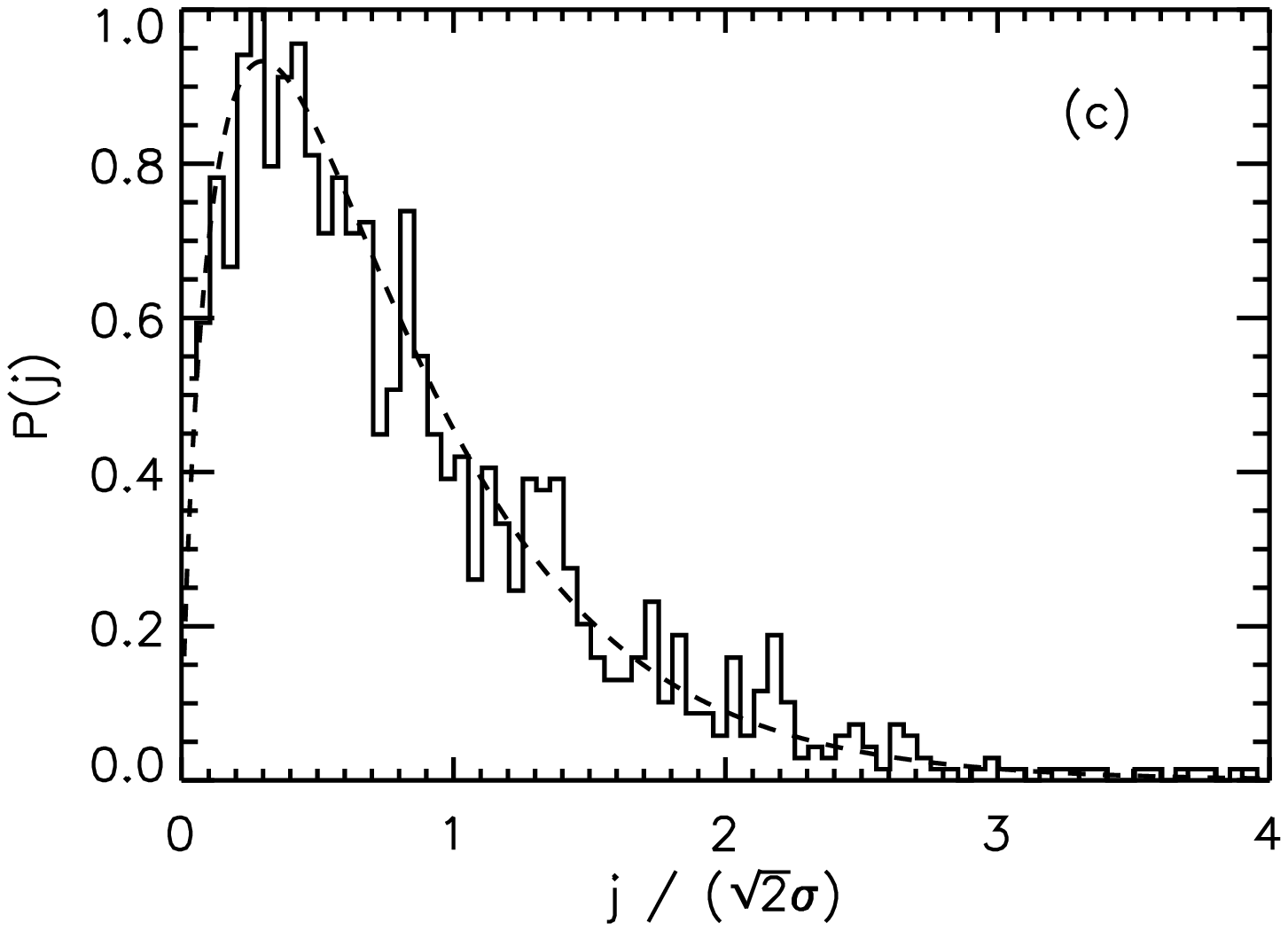}
\includegraphics[width=8cm]{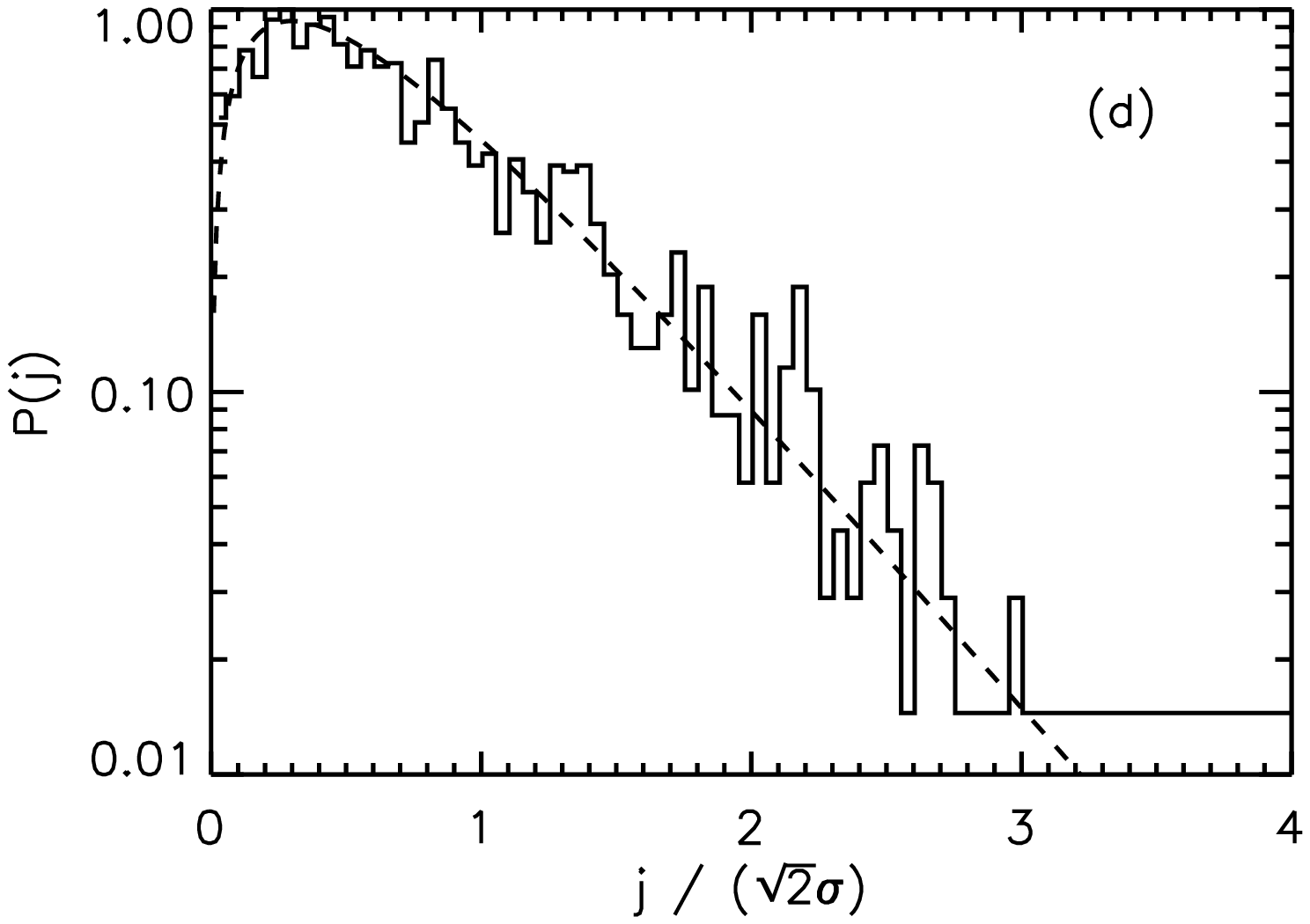}
\caption{Same as Fig.~\ref{fig:current_fb}, but for the open billiard at 5.77~GHz,
corresponding to Fig.~\ref{fig:billiards}c.}
\label{fig:current_qb}
\end{figure*}

In a chaotic billiard real and imaginary part of the wave function are uncorrelated within the random-super\-position-of-plane-waves approach,  $\langle\psi_R \psi_I\rangle=0$, where the average is taken over the billiard area. In a microwave measurement, there may be an additional rotation of the phase, caused by antennas, cables etc., i.e. the wave function registered by the apparatus is
\[
\tilde\psi = \tilde\psi_R + \imath\,\tilde\psi_I = e^{\imath\alpha} (\psi_R + \imath\,\psi_I).
\]
This has the consequence that real and imaginary part become correlated, $\langle\tilde\psi_R \tilde\psi_I\rangle \neq 0$. To begin with we have to remove this phase rotation. The procedure is described, e.g., in Refs. \cite{Ish01,Sai02}.

The distribution of intensities $\rho=|\psi|^2$ can be calculated with help of the Berry ansatz described above. If there is a complex mixing of independent Gaussian fields $\psi_R$ and $\psi_I$, we have in the transition regime \cite{Ish01,Sai02}
\begin{equation}\label{eq:prho}
P(\rho) = \mu \, \exp(-\mu^2\rho) \, I_0\!\left(\mu\sqrt{\mu^2-1}\rho\right)\,,
\end{equation}
where
\[
\mu = \frac{1}{2} \left( \epsilon + \frac{1}{\epsilon} \right)
\qquad\mathrm{and}\qquad
\epsilon = \sqrt{ \frac{\left\langle\psi_I^2\right\rangle}{\left\langle\psi_R^2\right\rangle} } \,.
\]

In the limit $\epsilon\to 0$, Eq.~(\ref{eq:prho}) describes the Porter-Thomas distribution found for systems where the wavefunction is real. For $\epsilon\to 1$, a single exponential behavior is observed which holds for systems where real and imaginary part are of the same strength.

Figures \ref{fig:intensity}a and b show the corresponding intensity
distributions for the ferrite billiard. The dashed lines are calculated from Eq.~(\ref{eq:prho}). The
parameter $\epsilon$ has not been fitted but has been taken directly from the experiment
by averaging $\psi_R^2$ and $\psi_I^2$ over all pixels at a given frequency.

One observes a quantitative agreement with theory for the case that the leakage current is negligible (Fig.~\ref{fig:intensity}b). 
For the scarred wavefunction of Fig.~\ref{fig:intensity}a, on the other hand, the experimentally found distribution of $|\psi|^2$ is completely at odds with theory. Such discrepancies are not new. Already in the disseminating paper by McDonald and Kaufman \cite{McD88} Gaussian distributions for $\psi$ were observed for chaotic wave functions exclusively. It is obvious that the random-superposition-of-plane-waves approach cannot work for bouncing-ball and scarred wave functions such as the one shown in Fig.~\ref{fig:billiards}a.
It is an easy matter to show that sharp drops in the $|\psi|^2$ distribution as the one in Fig.~\ref{fig:intensity}a are generic, e.g., for all wave functions of the rectangle \cite{Ish01}.

Figure \ref{fig:intensity}c shows a corresponding example for the open billiard at a frequency where
there is no leakage current. Again one finds a complete correspondence with theory.

\section{Current distributions}

For the currents shown in the bottom panel of Fig.~\ref{fig:billiards} the distribution function of 
\[
\vec{j} = \mathrm{\, Im}\left[\psi^*\nabla\psi\right] = \psi_R\nabla\psi_I-\psi_I\nabla\psi_R
\]
has to be calculated. A similar
problem occurred in our work on global and local level dynamics where the distributions
function of $\psi\nabla\psi$ was needed \cite{Bar99c}. With the
random-superposition-of-plane-waves approach the calculation of the averages is
straightforward and yields \cite{Sai02}
\begin{equation}\label{eq:pjxy}
P\left(j_{x,y}\right) = \frac{1}{\sqrt{2\langle j_{x,y}^2 \rangle}}\exp\left(-\sqrt{\frac{2}{\langle j_{x,y}^2 \rangle}}
\left|j_{x,y}\right|\right)
\end{equation}
for the distribution of the current components $j_x$ and $j_y$, and
\begin{equation}\label{eq:pj}
P(j) = \frac{4j}{\langle j^2 \rangle} K_0 \left( \frac{2j}{\sqrt{\langle j^2 \rangle}} \right)
\end{equation}
for the distribution of the modulus $j=\sqrt{j_x^2+j_y^2}$ where the parameter
\begin{equation}\label{eq:jparameter}
\langle j_{x,y}^2 \rangle = \frac{1}{2} \langle j^2 \rangle = k^2 \left<\psi_R^2\right>\left<\psi_I^2\right> 
\end{equation}
can again be taken directly from the experiment.

\begin{figure}
\includegraphics[width=8cm]{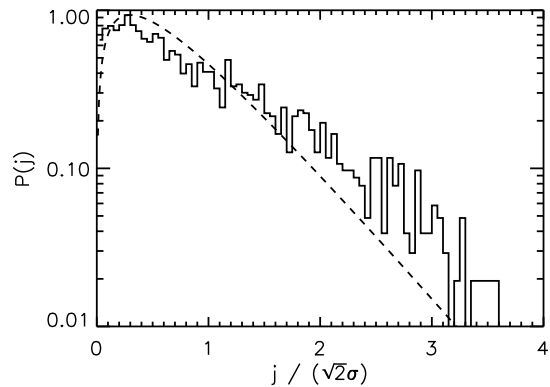}
\caption{Distribution of $j$ on a logarithmic scale as in Fig.~\ref{fig:current_fb}d, but for the scarred wave function at 5.19~GHz, shown in Fig.~\ref{fig:billiards}a.}
\label{fig:nocurrent}
\end{figure}

Figures \ref{fig:current_fb} and \ref{fig:current_qb} show the distributions for $j_x$, $j_y$, and $j$ for the
same patterns depicted in Figs. \ref{fig:billiards}b and c, respectively. Once more theory is in complete accordance with the experiment if there are no leakage currents, but is not able to describe the experiment otherwise. 
For comparison, Fig.~\ref{fig:nocurrent} shows the current distribution for the scarred wave function in Fig.~\ref{fig:billiards}a. Not unexpectedly there are again significant deviations from the universal behavior. The excess at large current values in particular is a consequence of the strong leakage currents observed at positions of high $|\psi|^2$ values (see Fig.~\ref{fig:billiards}a). 

In particular we observe identical distributions for $j_x$ and $j_y$ in the ferrite billiard. This is no longer the case for the open billiard, where the maximum of the $j_x$ distribution is shifted significantly to negative values caused by the transport from the right to the left through the billiard.

\section{The vortex spatial auto-correlation function}

A very useful quantity to characterize the vortex structure of a vector field is the
vorticity. Up to a factor 1/2 it is identical with the rotation of the current density,
\[
\omega = \frac{\partial\psi_R}{\partial x} \frac{\partial\psi_I}{\partial y}
  - \frac{\partial\psi_I}{\partial x} \frac{\partial\psi_R}{\partial y} =
  \frac{1}{2}\,\left(\nabla\times\vec{j}\right)_z\,.
\]

The calculation of the corresponding distribution function $P(\omega)$ follows exactly the same lines as for the current distribution and yields
\begin{equation}\label{eq:pomega}
P(\omega) = \frac{1}{\sqrt{2\langle\omega^2\rangle}} \exp\left( -\sqrt{\frac{2}{\langle\omega^2\rangle}}|\omega| \right),
\end{equation}
where 
\begin{equation}\label{eq:vparameter}
\langle\omega^2\rangle = \frac{1}{2} k^4 \langle\psi_R^2\rangle \langle\psi_I^2\rangle.
\end{equation}
can be taken directly from the experiment.

\begin{figure}
\includegraphics[width=8cm]{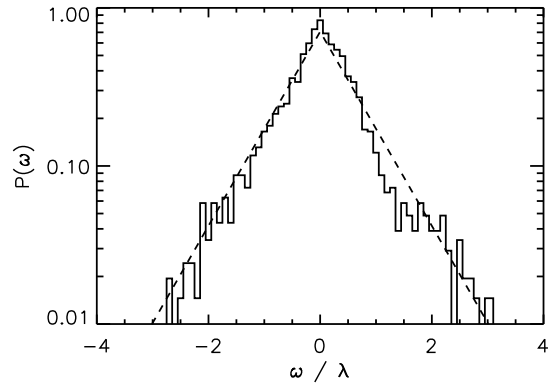}
\caption{Distribution of the vorticity for the ferrite billiard at 6.41~GHz corresponding to Fig.~\ref{fig:billiards}b. The dashed line corresponds to Eq.~(\ref{eq:pomega}), where $\lambda=\sqrt{\langle\omega^2\rangle}$, see Eq.~(\ref{eq:vparameter}).}
\label{fig:vorticity}
\end{figure}

Figure \ref{fig:vorticity} shows the vorticity distribution for the example shown in Fig.~\ref{fig:billiards}b for the ferrite billiard.
The dashed line has been calculated from Eq.~(\ref{eq:pomega}). Again the parameter $\lambda$
was not fitted but taken from the experiment. 

For the wave function to be zero both real and imaginary part have to be vanish. As a
consequence there are no longer nodal lines, but only nodal points. Each nodal point
corresponds to a vortex in the corresponding flow pattern. Since the distance between
neighboring node lines is of the order of half of a wave length both for real and
imaginary part, the mean spacing between neighboring nodal points is of this order of
magnitude as well.

Using the random-superposition-of-plane-waves approach, Berry and Dennis \cite{Ber00b} calculated two
types of vortex spatial auto-correlation functions, one of them called pair correlation
function and defined by
\begin{eqnarray}
  g(r) & = & g_0 \left\langle \delta[\psi_R(\bar{r}+r)]\,\delta[\psi_I(\bar{r}+r)]\,
  \delta[\psi_R(\bar{r})]\,\delta[\psi_I(\bar{r})]\, \right. \nonumber \\
  && \times 
  \left. |\omega(\bar{r}+r)|\,|\omega(\bar{r})| \right\rangle_{\bar{r}}\,. \label{eq:paircorr}
\end{eqnarray}
The normalization $g_0$ is determined such that $g(r)\to 1$ for $r\to\infty$. The other
correlation function discussed by the authors, the charge auto-correlation function
$g_Q(r)$, differs from expression (\ref{eq:paircorr}) only by the fact that the sign of the
vorticity is taken into account,
\begin{eqnarray*}
g_Q(r) & =& g_0 \left\langle \delta[\psi_R(\bar{r}+r)]\,\delta[\psi_I(\bar{r}+r)]\,
\delta[\psi_R(\bar{r})]\,\delta[\psi_I(\bar{r})]\, \right. \\
&& \times 
\left. \omega(\bar{r}+r)\,\omega(\bar{r})\right>_{\bar{r}}\,.
\end{eqnarray*}

Since for $g_Q(r)$ pairs of vortices with different senses of
rotation enter with a negative sign, we have  $g_Q(r)\to 0$ for $r\to\infty$.  For the
explicit expressions of $g(r)$ and $g_Q(r)$, which are quite complicated, the reader is
referred to the original work.
From the spatial auto-correlation function the distribution of nearest distances between
vortices can be calculated which has been studied by Saichev {\it et al.} \cite{Sai01}.

\begin{figure}
\includegraphics[width=8cm]{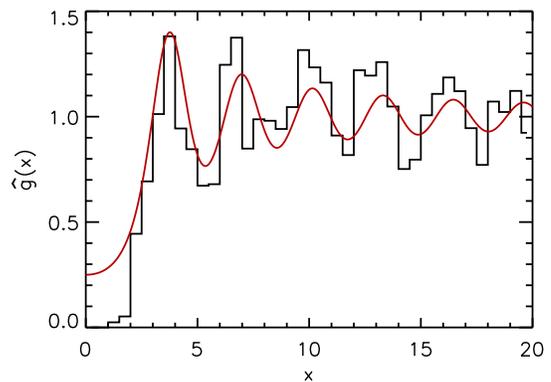}
\caption{Vortex pair correlation function $\hat{g}(x)$ for the ferrite billiard, see Eq.~(\ref{eq:paircorr}), where $x=kr$ and $\hat{g}(x)=g(r)$.
The solid line has been calculated from an integral expression given by Berry and Dennis \cite{Ber00b}.}
\label{fig:vortexcorr}
\end{figure}

The experimental study of the vortices is more difficult than that of the different
types of distribution functions discussed above, since it involves the determination of
the zeros of real and imaginary part of the wave function.
Because of the applied grid period of 5\,mm, the precision in determining distances is only moderate, and, more problematically, it is impossible to resolve vortices lying very close together which is very often the case for vortices with different sign.
In this situation a reliable determination of the distance distribution is not possible. The determination
of the charge correlation function $g_Q$, too, did not work because of the cancellation of
positive and negative terms, leading to an intolerable increase of the noise level. But
the pair correlation function $g$ could be determined. 
Since the wave number $k$ enters the pair correlation function as a scaling factor only, we may write $g(r)=\hat{g}(x)$, where $x=kr$. This allows to improve the statistics by superimposing the results from different frequencies. 
Figure \ref{fig:vortexcorr} shows the resulting pair correlation function $\hat{g}(x)$ obtained from current patterns at 43 different frequencies in the range from 4 to 8~GHz, by extracting the found vortices by hand. 
The frequency regimes showing a flow into the probe antenna were carefully avoided.
Though the statistics is only moderate, the oscillations predicted by theory are clearly observable. The hole
observed in the experimental histogram at small distances reflects the above mentioned difficulty to resolve closely neighboring vortices.

We can thus conclude that the experimentally obtained distributions of wave function
amplitudes, currents, and vortices are in quantitative agreement with the predictions of
the random-superposition-of-plane-waves approach, if the wavefunction is chaotic.

\begin{acknowledgments}
Discussions with K.-F. Berggren, Link\"oping, and with M. Vrani{\v c}ar, M. Robnik, Maribor, 
are gratefully acknowledged. The experiments were supported by the Deutsche
Forschungsgemeinschaft.
\end{acknowledgments}

\end{document}